\documentclass[final,1p,times]{elsarticle} 
\usepackage{graphicx} 
\usepackage{amssymb} 
\usepackage{amsthm} 
\usepackage{lineno} 

\journal{Nuclear Physics A} 
\begin{document} 

\begin{frontmatter} 


\title {Highlights from STAR: \\probing the early medium in heavy ion collisions}

\author{Gang Wang (for the STAR\footnote{For the full list of STAR authors and acknowledgements, see appendix `Collaborations' in this volume.} Collaboration)}

\address{University of California,
Los Angeles, California 90095, USA}
\ead{gwang@physics.ucla.edu}

\begin{abstract} 
We present highlights of recent results from the STAR Collaboration at RHIC,
focusing on the properties of the early medium created in heavy ion collisions.
We emphasize the strangeness production including the observation
of a hypernucleus (the hypertriton), the observation of 
reaction-plane-dependent angular correlation of charged particles
searching for local strong parity violation effects in heavy ion collisions,
and the evaluation of the medium viscosity from measurements of elliptic flow.
We discuss STAR's plan for the ``Critical Point Search" program at RHIC.
\end{abstract} 

\end{frontmatter} 



\section{Motivation}
In high-energy nuclear collisions at RHIC, 
a hot and dense matter has been shown to be formed with strong
collectivity developed at the partonic stage of the collisions~\cite{Whitepaper}.
This is an important step toward the understanding of the equation of state
of the medium created in such collisions. With its large acceptance and
excellent particle identification, the recent STAR program has been focusing on the study of the early medium,
such as strangeness production, collective motion, viscosity, and possible parity violation.
Strangeness is created in the collisions, and provides a special probe of the medium properties.
An enhanced strangeness observed in d+Au collisions (Sec. 2) improves
our understanding of the strangeness production mechanism~\cite{Zhu,Ant,Jinhui}.
In ultra-relativistic heavy ion collisions, metastable $\cal P$ and $\cal CP$
odd domains, characterized by non-zero topological charge, might be created~\cite{Lee}.
This {\it local strong parity violation}~\cite{Lee,Wick,Morley} can demonstrate itself via the asymmetry in the
emission of charged particle w.r.t. the system's angular momentum in non-central collisions.
The search for such charge separation will be discussed in Sec.~3~\cite{Sergei}.
The observation of the strong collectivity leads to the question if we have already reached the ideal hydrodynamical limit.
Recent STAR results on $v_2$ and $v_4$ anisotropic parameters show
deviation from ideal hydrodynamical values, indicating the presence of viscosity effects~\cite{Shusu,Patricia}.
We will show in Sec.~4 STAR's estimation of the shear viscosity over entropy density from
elliptic flow~\cite{Monika}.
This work will also make a case for the RHIC ``Critical Point Search"
program in Sec.~5 using the results based on the data successfully taken
at the lowest beam energies of $\sqrt{s_{NN}}$ = 9.2 GeV at RHIC~\cite{Lokesh,Tapan}.

\section{Strangeness production}
The enhancement of strangeness production in heavy ion collisions relative to that
in p+p collisions at the same energy were originally conceived
to be a smoking gun of Quark Gluon Plasma (QGP) formation \cite{Muller,STARprl}.
However, the strangeness enhancement in nuclear collisions relative to p+p could also be attributed
to the canonical suppression of strangeness production in p+p collisions \cite{canonical}.
The study of strangeness production in d+Au collisions will connect A+A and p+p data,
providing a better understanding of strangeness enhancement in nuclear collisions.

\begin{figure}
\center
\includegraphics[width=0.85\textwidth]{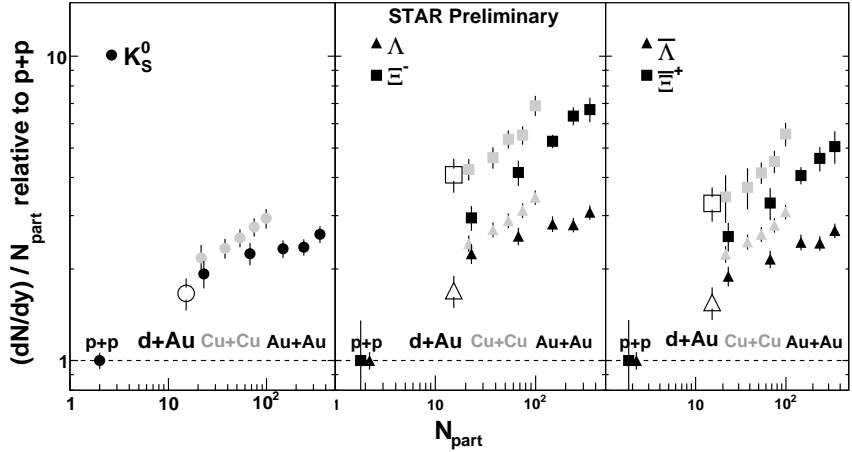}
\vspace{-0.3cm}
\caption{Strangeness enhancement factor ($dN/dy/N_{part}$ relative to p+p) for $K_{S}^{0}$ (left),
$\Lambda$, $\Xi^{-}$ (middle) and $\overline{\Lambda}$, $\overline{\Xi}^{+}$ (right)
from d+Au (open symbols), Cu+Cu (grey) and Au+Au (black) collisions at 200 GeV \cite{Zhu, Ant, Xiaobin}.}
\label{fig:dAu_enhance}
\end{figure}

\begin{figure}
\begin{minipage}[c]{0.48\textwidth}
\center
\includegraphics[width=\textwidth]{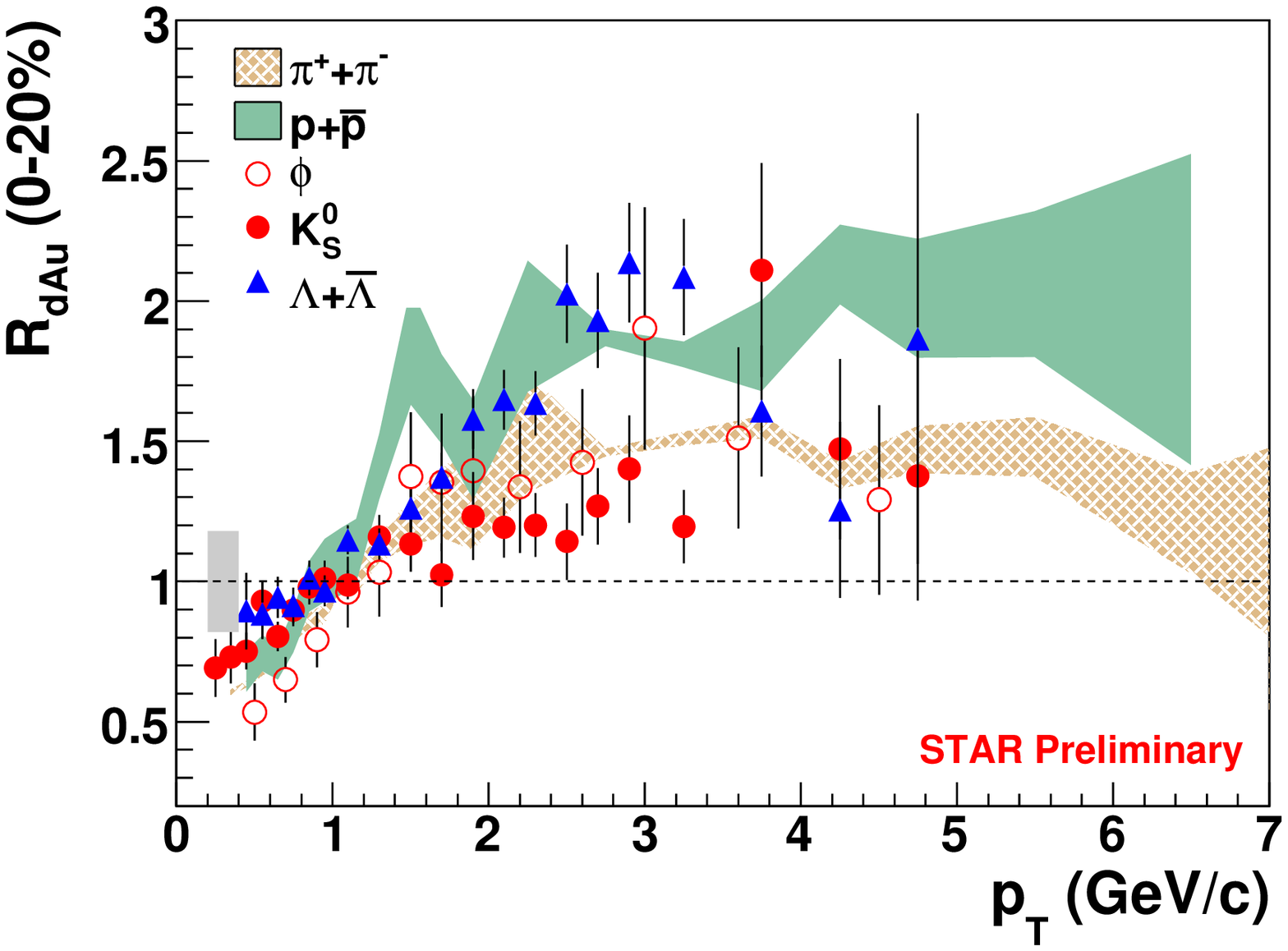}
  \caption{(color online) $R_{dAu}$ of strange hadrons ($K_{S}^{0}$, $\Lambda + \overline{\Lambda}$, $\phi$)
at mid-rapidity from 200 GeV d+Au collisions \cite{Zhu,phi}. $R_{dAu}$ of pions, kaons, protons
are from Ref.\cite{pkpi}.}
\label{fig:dAu_Raa}
\end{minipage}
\begin{minipage}[c]{0.48\textwidth}
\includegraphics[width=\textwidth]{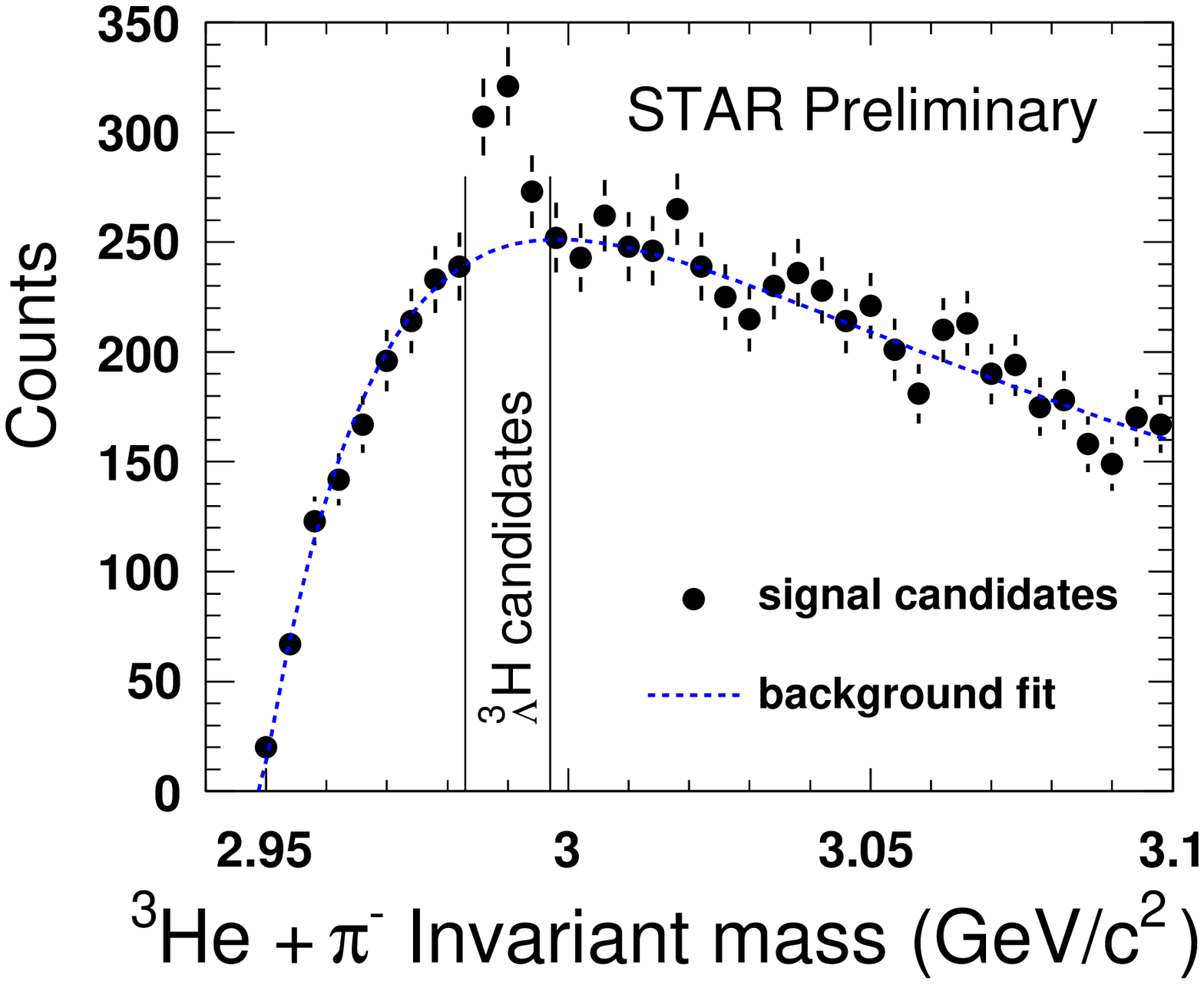}
  \caption{Preliminary result of the invariant mass distribution of $^{3}He+\pi^{-}$ at 200 GeV Au+Au collisions\cite{Jinhui}.
           The solid circles represent the signal candidate distribution, and the dashed line is the corresponding fitted
        background distribution.}
\label{fig:InvMass}
\end{minipage}
\end{figure}

The enhancement factor for a strange hadron can be defined as the yield per participant,
$N_{part}$, in heavy-ion collisions, divided by the respective value in p+p collisions at
the same center of mass energy per nucleon, and has a value of unity if strangeness production mechanism
in heavy-ion collisions is the same as in N-N collisions.
The enhancement factors are shown in Fig.~\ref{fig:dAu_enhance} as a function of $N_{part}$ for $K_{S}^{0}$,
$\Lambda$ and $\Xi$ from d+Au, Cu+Cu and Au+Au collisions at 200 GeV \cite{Zhu, Ant, Xiaobin}.
Fig.~\ref{fig:dAu_enhance} illustrates significant strangeness enhancement in the most central d+Au collisions,
especially for multi-strange hyperon $\Xi$. The enhancement factor is almost the same for
$K_{S}^{0}$ and $\Lambda$, and  much larger for $\Xi$, indicating that the enhancement factor increases with
the strangeness content. This observation seems to be consistent with the predictions from
the canonical statistical model \cite{canonical}, though the enhancement of strange hadrons 
could be not solely due to the canonical suppression in smaller systems~\cite{phi}.

The Cronin effect, the enhancement of hadron spectra at intermediate $p_T$
in p+A collisions with increasing nuclear size, was first seen in the nuclear modification
factor in low energy p+A collisions \cite{cronin}. Fig.~\ref{fig:dAu_Raa} shows that at RHIC energy
the nuclear modification factor in d+Au collisions demonstrates the Cronin effect for various particle species,
including $K_{S}^{0}$, $\Lambda + \overline{\Lambda}$ and $\phi$ \cite{Zhu,phi},
as well as pions and protons \cite{pkpi}.
The conventional models of the Cronin effect attribute the change of $p_T$ spectra in p+A to
multiple parton scattering in initial stage before hard parton scattering,
which will broaden the outgoing parton/hadron $p_T$ distribution \cite{scatter},
while the recombination model predicts the particle type (baryon/meson) dependence of the Cronin effect.
In Fig.~\ref{fig:dAu_Raa}, at intermediate $p_T$ ($2-3.5$ GeV/$c$), the nuclear modification factor $R_{dAu}$ is grouped into two bands:
baryons at about $1.81\pm 0.05$ and mesons around $1.28\pm 0.05$.
Note that the $\phi$ mass is close to the $\Lambda$ mass, but its $R_{dAu}$ is close to $K_S^0$'s.
This indicates the possible particle type dependence of $R_{dAu}$ instead of particle mass dependence.

A hypernucleus is a nucleus containing at least one hyperon.
The smallest and simplest hypernucleus is the hypertriton ($^{3}_{\Lambda}H$), consisting of a $\Lambda$, a proton and a neutron.
With a non-zero strangeness quantum number, the hypernucleus provides
one more degree of freedom for nuclear spectroscopy than a normal nucleus containing only protons and neutrons.
At RHIC energy, the hot and dense medium created in the collisions
liberates strange quarks with similar abundance to the up and down quarks,
leading to an abundance of nucleons, hyperons, mesons and their anti-particles,
some of which coalesce at a late stage of the collision and
produce nuclei and hypernuclei, or anti-nuclei and even anti-hypernuclei. 
In Fig.~3, we report the first observation of a hypernucleus at RHIC: the hypertriton $^{3}_{\Lambda}H$,
with a $5\sigma$ statistical significance,
via the invariant mass distribution of $^{3}He+\pi^{-}$ \cite{Jinhui}.
We also found that the production rate of $^{3}_{\Lambda}H / ^{3}He$ is close to unity~\cite{Jinhui},
which is different from the value obtained at the AGS, where an extra penalty factor has been observed 
with the appearance of the strangeness degree of freedom in particle production~\cite{AGS}. 
It seems that our results cannot be solely due to the finite size effect as implied by AGS data~\cite{AGS}, 
suggesting that strangeness phase space population is similar to light quarks at RHIC.

\section{Parity violation}
\begin{figure}
\begin{minipage}[r]{0.48\textwidth}
\includegraphics[width=\textwidth]{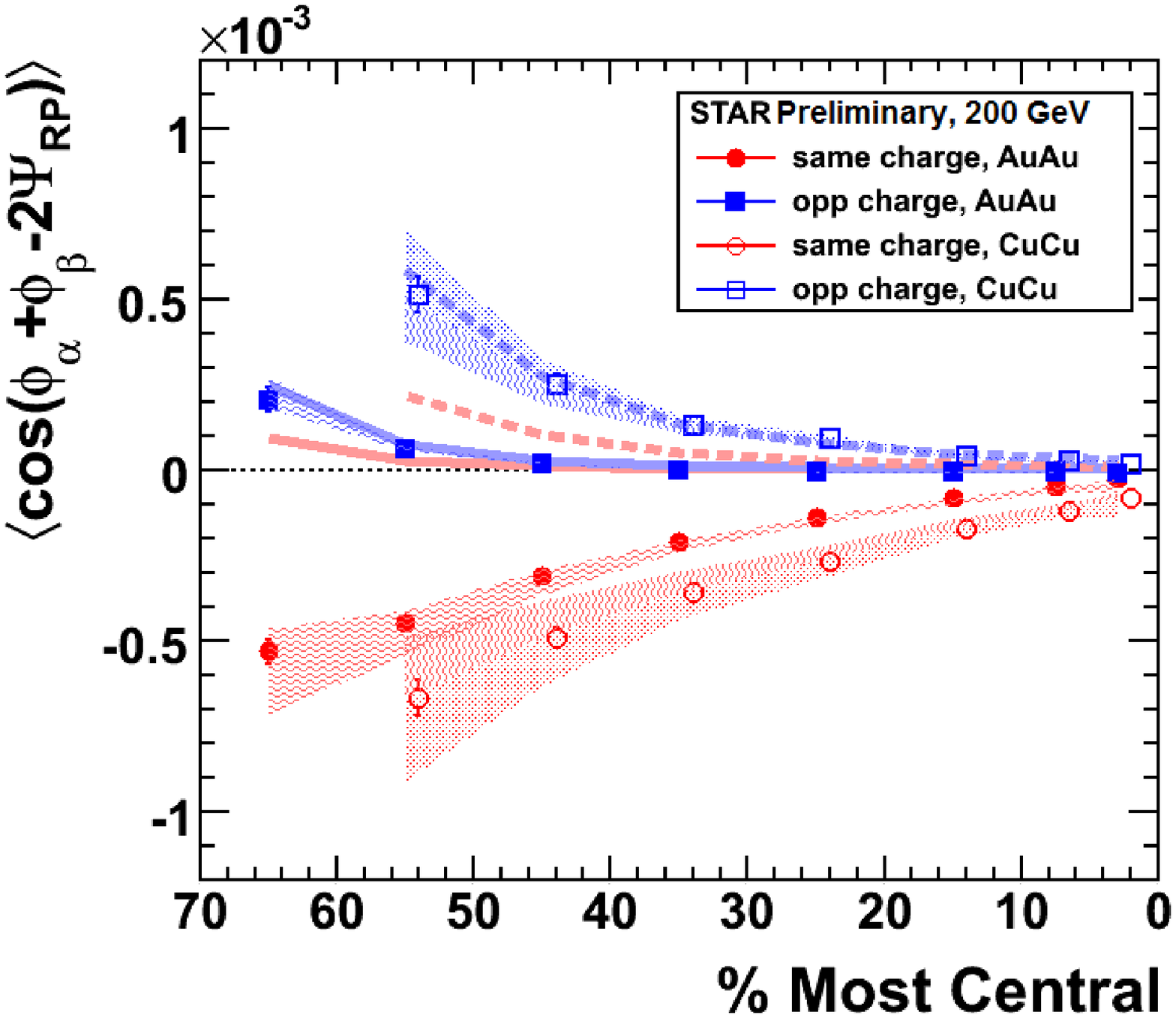}
  \caption{(color online) $\langle \cos(\phi_{\alpha} + \phi_{\beta} - 2\Psi_{RP}) \rangle$ in Au+Au and Cu+Cu collisions at 200 GeV~\cite{Sergei}.}
\label{fig:parity1}
\end{minipage}
\begin{minipage}[c]{0.48\textwidth}
\includegraphics[width=\textwidth]{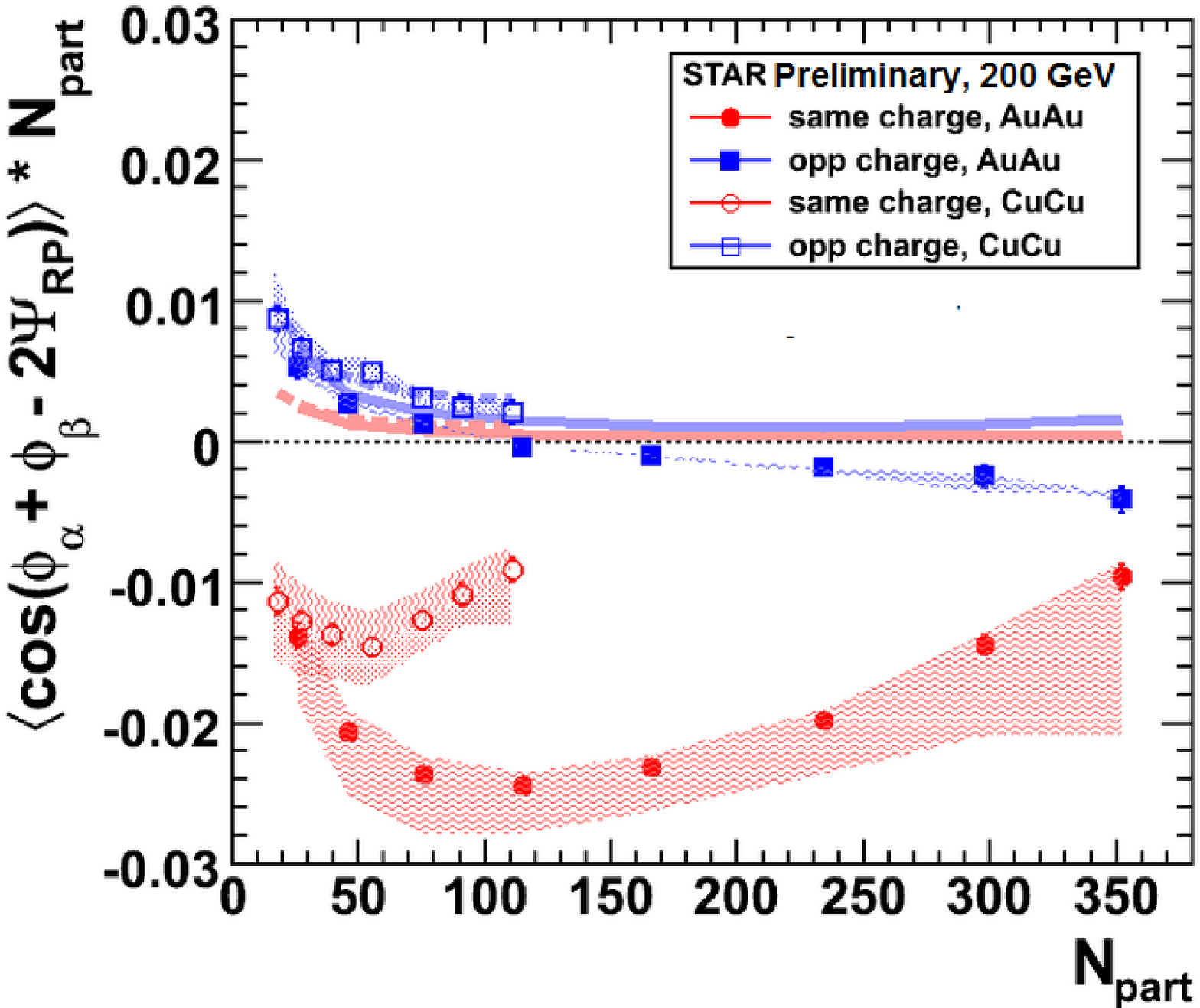}
  \caption{(color online) $\langle \cos(\phi_{\alpha} + \phi_{\beta} - 2\Psi_{RP}) \rangle$ scaled with the number of participants in Au+Au and Cu+Cu collisions at 200 GeV~\cite{Sergei}.}
\label{fig:parity2}
\end{minipage}
\end{figure}

The concept of local parity ($\cal P$) violation in high-energy heavy ion collisions
was discussed by Lee and Wick~\cite{Lee, Wick} and Morley and Schmidt~\cite{Morley} 
and elaborated by Kharzeev {\it et al.}~\cite{Dima}.
Recently it was argued~\cite{Dima2} that in non-central collisions such a $\cal P$-odd domain can manifest itself via preferential
same charge particle emission for particles moving along the system's angular momentum, due to the strong magnetic field produced
in the collision.
To study this effect, we investigate a three particle mixed harmonics azimuthal correlator,
$\langle \cos(\phi_{\alpha} + \phi_{\beta} - 2\Psi_{RP}) \rangle$~\cite{Sergei2}, which is
a {$\cal P$}-even observable, but can be sensitive to the charge separation effect.
In Fig.~\ref{fig:parity1}, we report the measurements of this observable using the STAR detector
in Au+Au and Cu+Cu collisions at 200 GeV~\cite{Sergei}. 
A negative value of the correlator for the
like-sign particle pairs means a positive correlation between them with respect to the reaction plane,
or that they tend to be emitted to the same side of the reaction plane. A positive value of the correlator
for the unlike-sign pairs means that they go to the opposite sides.

The correlations are weaker in more central collisions compared with more peripheral collisions,
which can be attributed partially to dilution of correlations in the case of particle production from multiple sources.
To compensate for this effect and to present a more complete picture of the centrality dependence, we show in
Fig.~\ref{fig:parity2} results multiplied by the number of participants ($N_{part}$)~\cite{Sergei}.
The decrease of the correlations in the most central collisions is expected as the magnetic field weakens.
The like- and unlike-sign correlations clearly exhibit very different behavior.
The unlike-sign correlations in Au+Au and Cu+Cu collisions are found to be very close at similar values of
$N_{part}$, supporting the picture of the correlator value being dominated
by the suppression of back-to-back correlations due to the medium~\cite{Back}.
The negative values of the unlike-sign correlator in central collisions may be caused by some non-${\cal P}$
effects, such as radial flow.
Existing event generators with known physics cannot describe our 
observation~\cite{Sergei}, and the local strong parity violation remains a possible explanation for the signal.

\section{Medium fluidity}
\begin{figure}
\center
\vspace{-0.3cm}
\includegraphics[width=0.85\textwidth]{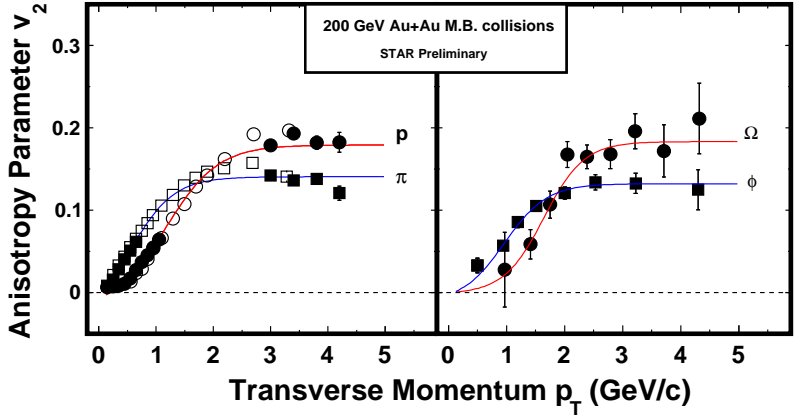}
\vspace{-0.3cm}
\caption{Preliminary results of $v_2(p_T)$ for $\pi$, $p$ (left) and $\phi$, $\Omega$ (right) in 200 GeV Au+Au collisions.
	Open and full symbols represent the results from PHENIX~\cite{PHENIX_v2} and STAR~\cite{Shusu}, respectively.
	Lines represent NQ-inspired fit.~\cite{Xin}}
\label{fig:v2_NQ}
\end{figure}

\begin{figure}[!btp]
\center
\vspace{-0.3cm}
\includegraphics[width=0.70\textwidth]{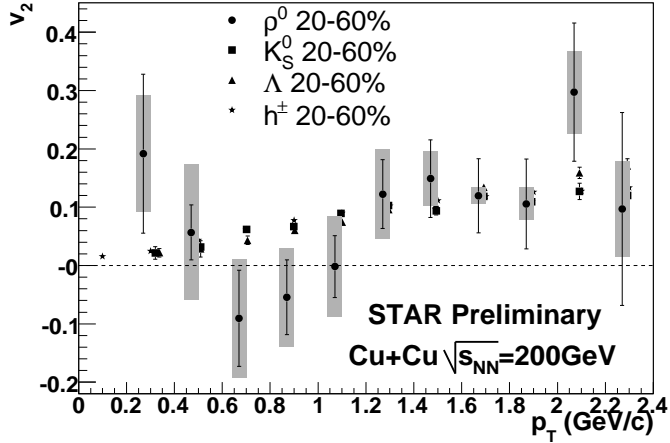}
\vspace{-0.3cm}
  \caption{Preliminary measurements of $\rho^{0} v_2$ in 200 GeV Cu+Cu collisions.~\cite{Patricia}}
\label{fig:Rho_v2}
\end{figure}

Measurements of elliptic flow ($v_2$) at RHIC demonstrate similar $p_T$-dependence
for $\pi$ ($p$) and $\phi$ ($\Omega$) as shown in Fig.~\ref{fig:v2_NQ}~\cite{PHENIX_v2,Shusu}.
This indicates that the heavier s quarks flow as strongly as    
the lighter u and d quarks, providing evidence for partonic collectivity.
$\rho^{0}$ vector meson is measured via its hadronic decay channel,
and the first measurement of $\rho^{0} v_2$ is shown in Fig.~\ref{fig:Rho_v2} 
for 200 GeV Cu+Cu collisions~\cite{Patricia}.
$\rho^{0}$ flows at $p_T > 1.2$ GeV/$c$, and we expect higher statistics in future RHIC runs
to explore the higher $p_T$ region to further test the partonic collectivity picture
and to further determine the production mechanism of $\rho^{0}$.

Combining the $v_2$ measurements at RHIC with ideal hydrodynamics calculations, we speculate that the hot and 
dense medium created in heavy ion collisions is a nearly perfect liquid~\cite{Whitepaper}.
To measure the medium fluidity, a dimensionless quantity, the ratio of shear viscosity to entropy density ($\eta / s$), 
is widely used.
Super-symmetric gauge theory~\cite{gauge} and uncertainty principle~\cite{uncertain}
render a conjectured lower bound for $\eta / s$ of $1/4\pi$. 

\begin{figure}
\center
\vspace{-0.3cm}
\includegraphics[width=0.70\textwidth]{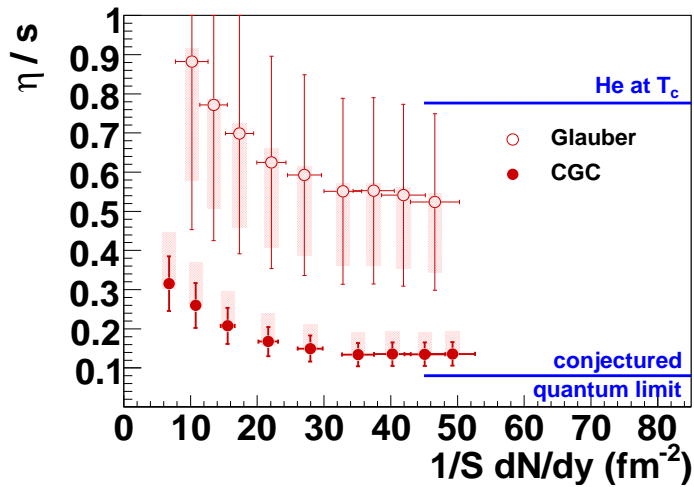}
\vspace{-0.3cm}
  \caption{(color online) $\eta / s$ as a function of $1/S dN/dy$ for 200 GeV Au+Au collisions. $v_2$ results used to determine $\sigma c_s$ are
from STAR preliminary (charged hadron), PHENIX (pion)~\cite{PHENIX_v2} and PHOBOS (charged hadron)~\cite{PHOBOS_v2}. See details in the text.}
\label{fig:etaOverS}
\end{figure}

Following Ref.~\cite{worlddata0}, if we consider the 
RHIC medium to be a classical ultrarelativistic gas of massless particles,
then $\eta / s = 0.316 \frac{T}{\bar{R} \sigma c_s} \frac{S}{dN/dy}$,
where the temperature  $T$ is obtained by fitting STAR's pion $m_T$ slope~\cite{mT},
$\sigma$ is the isotropic differential cross section of the massless particles, and $c_s$ is the speed of sound.
The product $\sigma c_s$ is determined by fitting $v_2/\epsilon$ with
$v_2/\epsilon = [v_2/\epsilon]_{hydro}\frac{1}{1+(\sigma c_s \frac{1}{S}dN/dy)/K_0}$~\cite{worlddata0}.
${\bar R}$, $S$ and $\epsilon$ are initial geometry parameters (mean
radius, area and eccentricity) of the collision system, 
estimated from Glauber (or CGC) calculations~\cite{Glauber,CGC}.
$K_0$ is a constant, determined through transport calculations to be 0.7~\cite{worlddata0}.
Fig.~\ref{fig:etaOverS} shows $\eta/s$ thus obtained as a function of $\frac{1}{S}dN/dy$,
the particle rapidity density per transverse area, for 200 GeV Au+Au collisions with two initial conditions (Glauber and CGC).
The $v_2$ results used to determine $\sigma c_s$ are
from STAR (charged hadron), PHENIX (pion)~\cite{PHENIX_v2} and PHOBOS (charged hadron)~\cite{PHOBOS_v2}.
Pions or charged hadrons are approximated to be massless particles in this approach.
The $\eta / s$ values from this approach are highly sensitive to the initial conditions,
but still suggesting that the RHIC medium indeed has a low viscosity, 
compared with another superfluid (Helium at its critical temperature).

An alternative technique to determine the medium viscosity was proposed by Gavin {\it et al.}~\cite{Gavin},
relying on the collision centrality evolution of two-particle $p_T$ correlation functions.
The $\eta/s$ value is extracted from the longitudinal broadening of the correlations with increasing collision centrality. 
The broadening arises from longitudinal diffusion of momentum currents, quantitatively determined by the magnitude 
of the kinematic viscosity, $\nu = \frac{\eta}{Ts}$, and the lifetime of the colliding system. 
We use differential extensions of the integral correlation observable $\tilde{C}$ proposed by Gavin {\it et al.}~\cite{Gavin}. 
The $\eta/s$ value thus obtained is found to be
highly sensitive to the freeze-out time estimate of the peripheral collisions,
which could impose large uncertainties to the estimated viscosity~\cite{Monika}.

\section{Critical Point Search}
Quantum Chromodynamics (QCD) critical point represents the end point of the first order phase transition boundary
in the phase diagram, spanned by the temperature and the baryonic chemical potential ($\mu_B$).
To search the QCD critical point and to explore the phase diagram, we experimentally vary $\mu_B$
and temperature by varying the beam energy.
As a first step of ``Critical Point Search" program at RHIC, a test
run was conducted in the year 2008 by colliding the Au ions at center of mass
energy 9.2 GeV. This short test run yielded about 3000 good events and
the results already show reasonably good statistical significance.

\begin{figure}
\begin{minipage}[r]{0.48\textwidth}
\includegraphics[width=\textwidth]{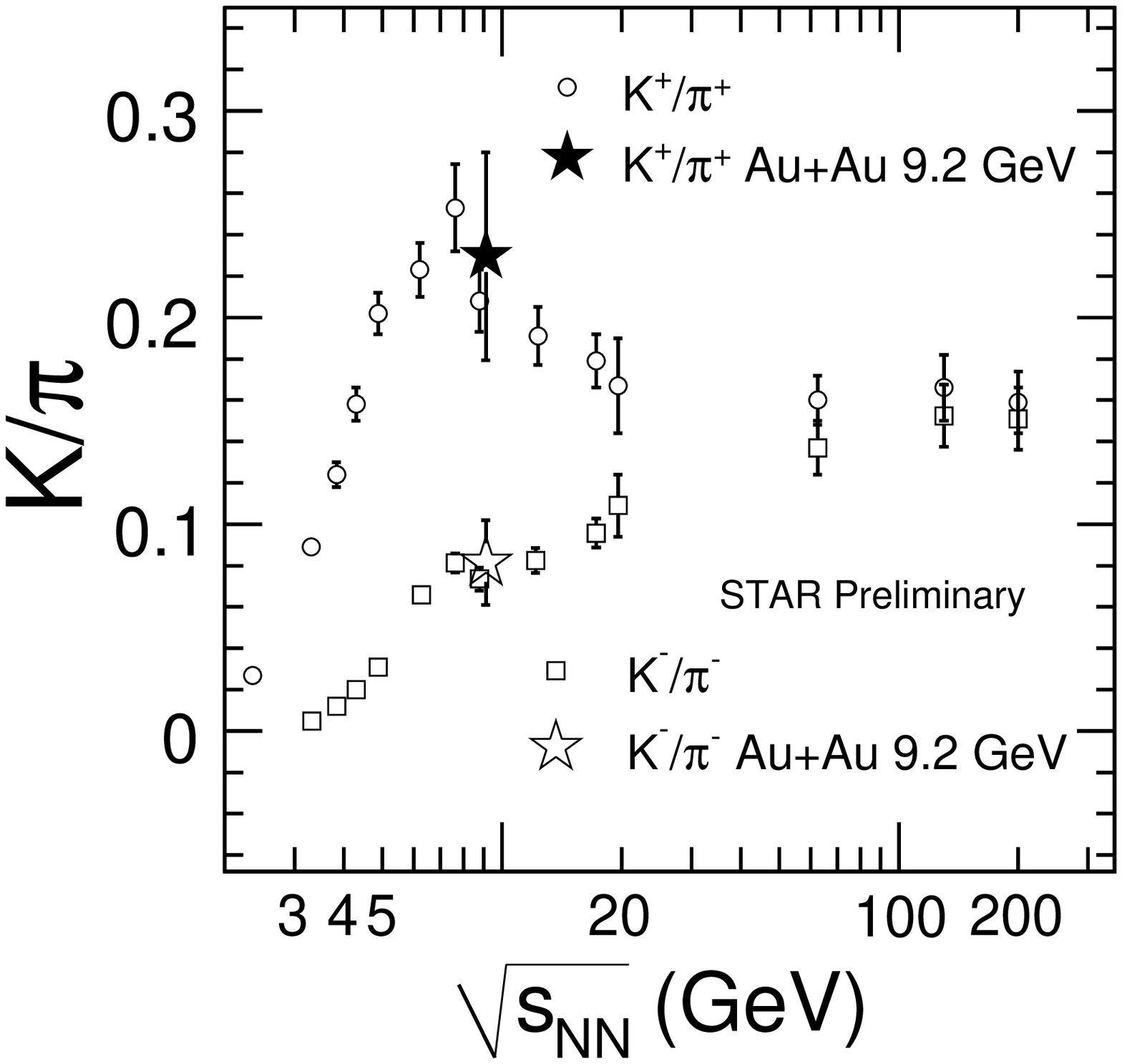}
  \caption{$K/\pi$ plotted as a function of beam energy \cite{Lokesh,SPS}.}
\label{fig:k2pi_9GeV}
\end{minipage}
\begin{minipage}[c]{0.48\textwidth}
\includegraphics[width=\textwidth]{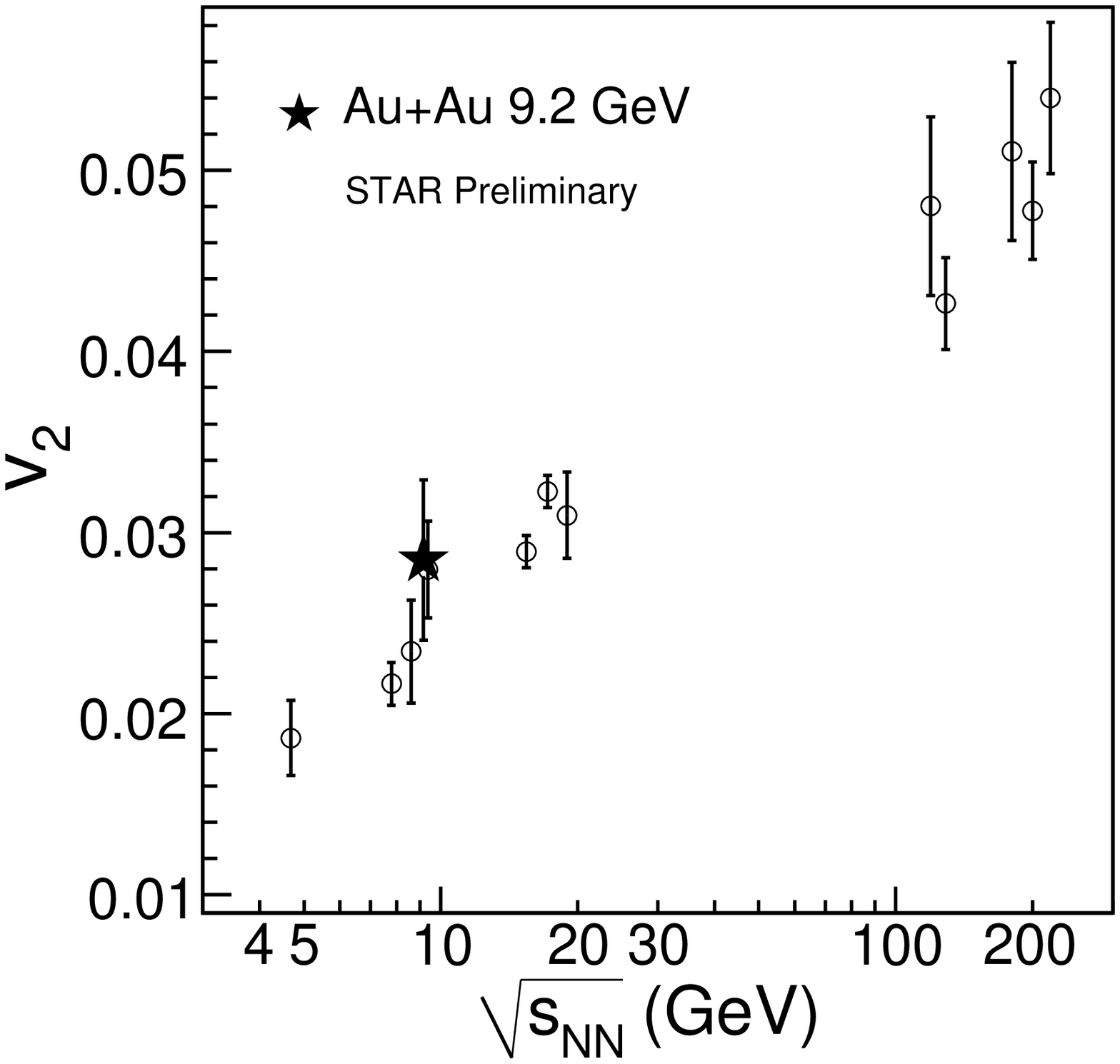}
  \caption{Energy dependence of charged hadron $v_2$ near mid-rapidity ($-1 < \eta< 1$) \cite{Lokesh,STAR,E877,NA49,PHENIX,PHOBOS}.}
\label{fig:v2_9GeV}
\end{minipage}
\end{figure}

Fig.~\ref{fig:k2pi_9GeV} and Fig.~\ref{fig:v2_9GeV} show respectively the particle ratio $K/\pi$ and charged hadron $v_2$
at 9.2 GeV Au+Au collisions~\cite{Lokesh}, following the trend from other center of mass
energies~\cite{SPS,STAR,E877,NA49,PHENIX,PHOBOS}.
As a collider experiment, STAR has many advantages over the fixed target experiments in terms of
acceptance ($p_T$, $\eta$), particle density per unit area at a fixed $\sqrt{s_{NN}}$
and the particle identification. The latter will be further improved by the inclusion of Time Of Flight (TOF)~\cite{TOF},
and will largely benefit analyses such as particle ratio fluctuations.
The results in Au+Au 9.2 GeV have demonstrated that
STAR is ready both technically and scientifically to pursue the future ``Critical Point Search" program.

One of the characteristic signatures of the critical point is an increase in fluctuations of various 
event-by-event observables~\cite{Stephanov}.
The moments of the conserved quantities such as net charge and net-baryon are related to respective 
thermodynamic quantity susceptibilities.
We have carried out a study of mean, standard deviation, skewness and kurtosis of
event-by-event net-charge and net-proton distributions in 200 GeV Au+Au and Cu+Cu
collisions~\cite{Tapan},
helping to understand the expectations from various physics processes at different collision energies.

\section{Summary}
We have reported STAR's most recent progress towards understanding the properties
of the early medium at RHIC. The strangeness enhancement measurements in d+Au collisions
favor the canonical statistical model~\cite{canonical}, and the corresponding $R_{dAu}$ results support
the recombination model. For the first time at RHIC, we have observed a hypernucleus
(the hypertriton) with a $5\sigma$ statistical significance.
Qualitatively the charge-separation results agree with the general features
of the theoretical predictions for local ${\cal P}$-violation in heavy ion collisions.
For 200 GeV Au+Au collisions, medium shear viscosity over entropy density has been estimated from elliptic flow. 
We have demonstrated STAR's readiness for the ``Critical Point Search" program with our test run physics results.


\end{document}